\documentclass[a4paper]{jpconf}

\usepackage{graphicx}

\usepackage{bm}% bold math
\usepackage{color} % color fonts

\newcommand{\vr}{\bm{r}}

\begin{document}

\title{Neutron halo in deformed nuclei from a relativistic Hartree-Bogoliubov model in a Woods-Saxon basis}

\author{S G Zhou$^{1,2}$, J Meng$^{3,1,2}$, P Ring$^{4,3}$ and E G Zhao$^{1,2,3}$}
 \address{$^1$ Institute of Theoretical Physics, Chinese Academy of Sciences,
              Beijing 100190, China}
 \address{$^2$ Center of Theoretical Nuclear Physics, National Laboratory
              of Heavy Ion Accelerator, Lanzhou 730000, China}
 \address{$^3$ School of Physics, Peking University,
              Beijing 100871, China}
 \address{$^4$ Physikdepartment, Technische Universit\"at M\"unchen,
              85748 Garching, Germany}
 \ead{sgzhou@itp.ac.cn}

\begin{abstract}
Halo phenomenon in deformed nuclei is studied by using a fully
self-consistent deformed relativistic Hartree-Bogoliubov model in a
spherical Woods-Saxon basis with the proper asymptotic behavior at
large distance from the nuclear center. Taking a deformed
neutron-rich and weakly bound nucleus $^{44}$Mg as an example and by
examining contributions of the halo, deformation effects, and large
spatial extensions, we show a decoupling of the halo orbitals from
the deformation of the core.
\end{abstract}

\section{Introduction}

Since it was first observed in the weakly bound system
$^{11}$Li~\cite{Tanihata1985}, halo phenomenon has been one of the
most interesting topics in nuclear physics. Much effort has been
focused on the investigation of the structure and dynamics of nuclear
halo~\cite{Jensen2004}. Since most open shell nuclei are deformed,
the interplay between deformation and weak binding raises interesting
questions, such as whether or not there exist halos in deformed
nuclei and, if yes, what are their new features.

Calculations in a deformed single-particle model with the spin-orbit
coupling neglected have shown that valence particles in specific
orbitals with low projection of the angular momentum on the symmetry
axis can give rise to halo structures in the limit of weak binding
and the deformation of the halo may be different from that of the
core~\cite{Misu1997}. Halos in deformed nuclei were investigated in
several mean field calculations~\cite{Li1996, Pei2006, Nakada2008}.
However, there are some doubt about the occurrence of halos in
deformed nuclei. For example, it has been concluded that in the
neutron orbitals of an axially deformed Woods-Saxon potential the
lowest-$\ell$ component becomes dominant at large distances from the
origin and therefore all $\Omega^{\pi} = 1/2^+$ levels do not
contribute to deformation for binding energies close to
zero~\cite{Hamamoto2004}. In addition, a three-body model
study~\cite{Nunes2005} suggests that it is unlikely to find halos in
deformed drip line nuclei because the correlations between the
nucleons and those due to static or dynamic deformations of the core
inhibit the formation of halos.

In order to give an adequate description of halos in deformed
nuclei, a model should be used which includes in a self-consistent
way the continuum, deformation effects, large spatial distributions,
and couplings among all these features. Spherical nuclei with halos
have been described in the past successfully by the solution of
either the non-relativistic Hartree-Fock-Bogoliubov
(HFB)~\cite{Bulgac1980, Dobaczewski1984, Dobaczewski1996} or the
relativistic Hartree Bogoliubov (RHB) equations ~\cite{Meng1996,
Poschl1997, Meng1998b} in coordinate ($r$) space. However, for
deformed nuclei the solution of HFB or RHB equations in $r$ space is
a numerically very demanding task. In the past considerable effort
has been made to develop mean field models either in $r$ space or in
a basis with an improved asymptotic behavior at large
distances~\cite{Nakada2008, Terasaki1996, Stoitsov2000, Teran2003,
Zhou2003a, Tajima2004, Stoitsov2008, Pei2008}. In
Ref.~\cite{Zhou2003a} the Woods-Saxon basis was proposed as a
reconciler between the harmonic oscillator basis and coordinate
space. The Woods-Saxon wave functions have more realistic asymptotic
behavior at large $r$ than the harmonic oscillator wave functions
do. One can use a box boundary condition to discretize the
continuum. It has been shown that the results in a Woods-Saxon basis
is almost equivalent to those obtained in $r$ space~\cite{Zhou2003a,
Schunck2008a, Schunck2008}. A deformed relativistic Hartree model
(DRH)~\cite{Zhou2006} and a deformed relativistic Hartree-Bogoliubov
model (DRHB)~\cite{Zhou2008a} in a Woods-Saxon basis have also been
developed.

In a recent work~\cite{Zhou2010}, the halo phenomenon in deformed
nuclei is studied by using the DRHB model in a Woods-Saxon basis. In
this contribution, we shall present some of the results on neutron
halo in deformed nuclei. The formalism of the DRHB model in a
Woods-Saxon basis will be given in section~\ref{sec:formalism}. In
section~\ref{sec:results}, the results and discussions will be
presented. Finally a summary is given.

\section{\label{sec:formalism}The deformed relativistic Hartree-Bogoliubov model in a Woods-Saxon basis}

The RHB equation for the nucleons reads~\cite{Kucharek1991}
%\begin{widetext}
\begin{eqnarray}
% &   &
% \hspace*{-0.4cm}
 \sum_{\sigma'p'} \int d^3 \vr'
 \left(
  \begin{array}{cc}
   h_\mathrm{D}(\vr\sigma p,\vr\sigma'p') - \lambda &
   \Delta(\vr\sigma p,\vr'\sigma' p') \\
  -\Delta^*(\vr\sigma p,\vr'\sigma' p')
   & -h_\mathrm{D}(\vr\sigma p,\vr\sigma'p') + \lambda \\
  \end{array}
 \right)
% \nonumber \\
% &   & \mbox{} \times
 \left(
  { U_{k}(\vr'\sigma' p') \atop V_{k}(\vr'\sigma' p') }
 \right)
 & = &
 E_{k}
  \left(
   { U_{k}(\vr\sigma p) \atop V_{k}(\vr\sigma p) }
  \right)
 ,
 \nonumber \\
 \label{eq:RHB0}
\end{eqnarray}
%\end{widetext}
where $p=1, 2$ is used to represent the particle-antiparticle degree
of freedom, $E_{k}$ is the quasiparticle energy, $\lambda$ is the
Fermi energy, and $h_\mathrm{D}$ is the Dirac
Hamiltonian~\cite{Serot1986, Reinhard1989, Ring1996, Vretenar2005,
Meng2006},
\begin{equation}
 h_\mathrm{D} =
  \bm{\alpha} \cdot \bm{p} + V(\bm{r}) + \beta (M + S(\bm{r}))
  .
\label{eq:Dirac0}
\end{equation}
The pairing potential reads
\begin{eqnarray}
 \Delta(\vr_1\sigma_1 p_1,\vr_2\sigma_2 p_2)
 & = & %\frac{1}{2}
   \sum^{\sigma'_2p'_2}_{\sigma'_1p'_1}
   V_{p_1p_2p'_1p'_2}(\vr_1,\vr_2;\sigma_1\sigma_2\sigma'_1\sigma'_2)
% \nonumber \\
% &  & \mbox{}
%   \times
   \kappa(\vr_1\sigma'_1 p'_1,\vr_2\sigma_2' p_2')
 \ .
\end{eqnarray}

For axially deformed nuclei with spacial reflection symmetry, we
expand the potentials and the densities in terms of the Legendre
polynomials~\cite{Price1987},
\begin{equation}
 f(\bm{r})   = \sum_\lambda f_\lambda({r}) P_\lambda(\cos\theta),\
 \lambda = 0,2,4,\cdots.
 \label{eq:expansion}
\end{equation}
The quasiparticle wave function is expanded in terms of wave
functions of the Dirac Woods-Saxon basis $\left\{ \epsilon_{i\kappa
m}, \varphi_{i\kappa m}(\bm{r}\sigma p) \right\}$ as,
\begin{equation}
   U_{k} (\vr\sigma p)
  = \displaystyle
  \sum_{i\kappa}
  \left(
  { u^{(m)}_{k,(i\kappa)} \varphi_{i\kappa m}(\vr\sigma p)
     \atop
     u^{(\bar m)}_{k,(\widetilde{i\kappa})} \tilde\varphi_{i\kappa m}(\vr\sigma p)
   }
  \right)
 \ , \
% \label{eq:Uexpansion0}
%\end{equation}
%\begin{equation}
   V_{k} (\vr\sigma p)
  = \displaystyle
  \sum_{i\kappa}
  \left(
   {
     v^{(m)}_{k,({i\kappa})} \varphi_{i\kappa m}(\vr\sigma p)
     \atop
     v^{(\bar m)}_{k,(\widetilde{i\kappa})} \tilde\varphi_{i\kappa m}(\vr\sigma p)
   }
  \right)
 .
 \label{eq:Vexpansion0}
\end{equation}
The basis wave function reads
\begin{equation}
 \varphi_{i\kappa m}(\vr\sigma) =
   \frac{1}{r}
   \left(
     \begin{array}{c}
       i G_{i\kappa}(r) Y^l _{jm} (\Omega\sigma)
       \\
       - F_{i\kappa}(r) Y^{\tilde l}_{jm}(\Omega\sigma)
     \end{array}
   \right) ,
   \ \ j = l\pm\frac{1}{2},
% \label{eq:SRHspinor}
\end{equation}
with $G_{i\kappa}(r) / r$ and $F_{i\kappa}(r) / r$ the radial wave
functions for the upper and lower components and $Y^l _{jm}$ the
spinor spherical harmonics where $\kappa = (-1)^{j+l+1/2} (j+1/2)$
and $\tilde l = l + (-1)^{j+l-1/2}$. $\tilde\varphi_{i\kappa
m}(\vr\sigma p)$ is the time reversal state of $\varphi_{i\kappa
m}(\vr\sigma p)$. The states both in the Fermi sea and in the Dirac
sea should be included in the basis for the
completeness~\cite{Zhou2003a, Zhou2003}. For each $m$-block, solving
the RHB equation (\ref{eq:RHB0}) is equivalent to the diagonalization
of the matrix,
\begin{equation}
 \left( \begin{array}{cc}
  {\cal A} & {\cal B} \\
  {\cal C} & {\cal D} \\
 \end{array} \right)
 \left(
  { {\cal U}
    \atop
    {\cal V}
  }
 \right)
 = E
 \left(
  { {\cal U}
    \atop
    {\cal V}
  }
 \right),
\end{equation}
where
\begin{equation}
 {\cal U} = \left(u^{(m)}_{k,(i\kappa)}\right),\
 {\cal V} = \left(v^{(m)}_{k,(\widetilde{i\kappa})}\right),
\end{equation}
\begin{equation}
 {\cal A} = \left(A^{(m)}_{(i\kappa)(i'\kappa')}\right),\
 {\cal D} =
  \left(-A^{(m)}_{(\widetilde{i\kappa})(\widetilde{i'\kappa'})}\right),
\end{equation}
\begin{equation}
 {\cal B} =
  \left(\Delta^{(m)}_{(i\kappa)(\widetilde{i'\kappa'})}\right),\
 {\cal C} =
  \left(-\Delta^{(m)}_{(\widetilde{i\kappa})(i'\kappa')}
 =\Delta^{(m)}_{(i'\kappa')(\widetilde{i\kappa})} \right).
 \label{eq:pairing_matrix}
\end{equation}

For the pp channel, we use a zero range density dependent force,
%\begin{widetext}
\begin{equation}
 V_{p_1p_2p'_1p'_2}(\vr_1,\vr_2;\sigma_1\sigma_2\sigma'_1\sigma'_2)
 = \frac{1}{4} V_0 \delta( \vr_1 - \vr_2 )
   \left(1-\frac{\rho(\vr_1)}{\rho_\mathrm{sat}}\right)
   \left[ 1 - 4 \vec{\sigma}_{11'} \cdot \vec{\sigma}_{22'} \right]
   \left[ \mathbf{I}^p_{11'} \cdot \mathbf{I}^p_{22'} \right]
 .
 \label{eq:pairing_force}
\end{equation}
%\end{widetext}

\section{\label{sec:results}Results and discussions}

\begin{figure}
\includegraphics[width=10cm]{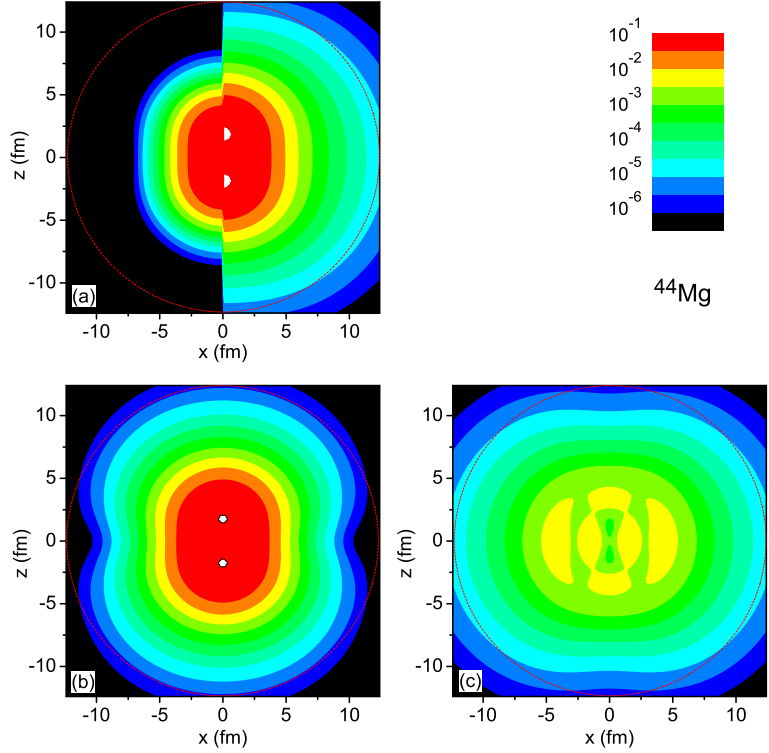}\hspace{2pc}%
\begin{minipage}[b]{12pc}
\caption{\label{fig1}%
(Color online) Density distributions of $^{44}$Mg with the $z$-axis
as symmetry axis: (a) the proton density (for $x<0$) and the neutron
density (for $x>0$), (b) the density of the neutron core, and (c)
the density of the neutron halo. In each plot, a dotted circle is
drawn for guiding the eye. This figure is originally published in Ref.~\cite{Zhou2010}.}
\end{minipage}
\end{figure}

The calculations are based on the density functional
NL3~\cite{Lalazissis1997}. For the pp
interaction~(\ref{eq:pairing_force}), the following parameters are
used: $\rho_\mathrm{sat} =$ 0.152~fm$^{-3}$ and $V_0 =
380$~MeV$\cdot$fm$^3$, and a cut-off energy
$E^\mathrm{q.p.}_\mathrm{cut} = 60$~MeV is applied in the
quasi-particle space. These parameters were fixed by reproducing the
proton pairing energy of the spherical nucleus $^{20}$Mg obtained
from a spherical relativistic Hartree-Bogoliubov calculation with the
Gogny force D1S. A spherical box of the size $R_\mathrm{max} = 20$ fm
and the mesh size $\Delta r = 0.1$ fm are used for generating the
spherical Dirac Woods-Saxon basis~\cite{Zhou2003a} which consists of
states with $j< \frac{21}{2} \hbar$. An energy cutoff
$E^+_\mathrm{cut}$ = 100 MeV is applied to truncate the positive
energy states in the Woods-Saxon basis and the number of negative
energy states in the Dirac sea is taken to be the same as that of
positive energy states in each ($\ell,j$)-block.

In the DRHB calculations for magnesium isotopes, the last nucleus
within the neutron drip-line is $^{46}$Mg which is almost spherical.
The neighboring even-even nucleus $^{44}$Mg is well deformed with
quadrupole deformation $\beta_2 = 0.32$. This nucleus is weakly
bound with a small two-neutron separation energy $S_{2n} = 0.44$
MeV. Since we are interested in the neutron halo in deformed
unstable nuclei, $^{44}$Mg is taken as an example for a detailed
investigation. The density distributions of all protons and all
neutrons in this nucleus are shown in Figure~\ref{fig1}(a). Due to
the large neutron excess, the neutron density not only extends much
farther in space but also shows a halo structure. The neutron
density is decomposed into the contribution of the core in
Figure~\ref{fig1}(b) and that of the halo in Figure~\ref{fig1}(c).
Details of this decomposition are given later. It is seen that the
core of $^{44}$Mg is prolately deformed, but the halo has a slightly
oblate deformation, which indicates the decoupling between the
deformations of core and halo.

\begin{figure}
\includegraphics[width=9.0cm]{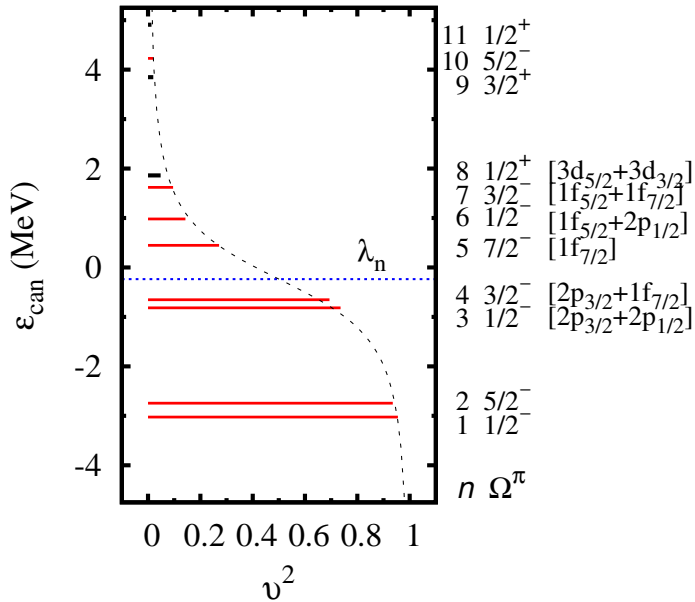}\hspace{2pc}%
\begin{minipage}[b]{14pc}
\caption{\label{fig2} (Color online) Single neutron levels with the
quantum numbers $\Omega^\pi$ around the chemical potential (dotted
line) in the canonical basis for $^{44}$Mg as a function of the
occupation probability $v^2$. The order $n$, $\Omega^\pi$, and the
main Woods-Saxon components for orbitals close to the threshold are also
given. The dashed line corresponds to the BCS-formula with an average
pairing gap. This figure is originally published in Ref.~\cite{Zhou2010}.}
\end{minipage}
\end{figure}

To study the formation mechanism of a nuclear halo, one needs to
investigate the weakly bound orbitals and/or those embedded in the
continuum. For an intuitive understanding of the single particle
structure we keep in mind that HB-wave functions can be represented
by BCS-wave functions in the canonical basis and show in
Figure~\ref{fig2} the corresponding single particle spectrum for the
neutrons. As discussed in Ref.~\cite{Ring1980} the single particle
energies in the canonical basis $\varepsilon_k=\langle
k|h_D|k\rangle$ shown in Figure~\ref{fig2} are expectation values of
the Dirac Hamiltonian~(\ref{eq:Dirac0}) for the eigenstates
$|k\rangle$ of the single particle density matrix $\hat\rho$ with the
eigenvalues $v^2_k$. The discrete part of the spectrum of $\hat\rho$
with $v^2_k>0$ contributes to the HB-wave function and only this part
is plotted in Figure~\ref{fig2}. This part of the spectrum
$\varepsilon_k$ is discrete even for the levels in the continuum. Of
course, this is only possible because the wave functions $|k\rangle$
are not eigenfunctions of the Hamiltonian. As long as the Fermi
energy $\lambda_\mathrm{n}$ is negative, the corresponding density
$\rho(\vr)$ is localized and the particles occupying the levels in
the continuum are bound~\cite{Dobaczewski1984}.

The orbitals in Figure~\ref{fig2} are labeled by the conserved
quantum numbers $\Omega$ and $\pi$. The character $n$ numbers the
different orbitals appearing from the bottom to the top in this
figure according to their energies. The neutron Fermi energy lies
within the $pf$ shell and most of the single particle levels have
negative parities. Since the chemical potential $\lambda_\mathrm{n}
= -230$ keV is relatively small, the orbitals above the threshold
have noticeable occupation probabilities due to pairing
correlations. For example, the occupation probabilities of the 5th
($\Omega^\pi = 7/2^-$) and the 6th ($\Omega^\pi = 1/2^-$) orbitals
are 27.2\% and 14.3\%, respectively.

As we see in Figure~\ref{fig2} there is a considerable gap between
the two levels with the numbers $n=2$ and $n=3$. The levels with
$\varepsilon_\mathrm{can} < -2.5$ MeV contribute to the ``core'',
and the other remaining weakly bound and continuum orbitals with
$\varepsilon_\mathrm{can} > -1$ MeV naturally form the ``halo''.
Therefore we decompose the neutron density into two parts, one part
coming from the orbitals with canonical single particle energies
$\varepsilon_\mathrm{can} < -2.5$ MeV (called ``core'') and the
other from the remaining weakly bound and continuum orbitals (called
``halo''). A further decomposition of the neutron density shows that
the two weakly bound orbitals, i.e., the 3rd ($\Omega^\pi = 1/2^-$)
and the 4th ($\Omega^\pi = 3/2^-$), contribute mostly to the halo.
If we decompose the deformed wave functions of the two weakly bound
orbitals, i.e. the 3rd ($\Omega^\pi = 1/2^-$) and the 4th
($\Omega^\pi = 3/2^-$), in the spherical Woods-Saxon basis it turns
out that in both cases the major part comes from $p$ waves as
indicated on the right hand side of Figure~\ref{fig2}. The low
centrifugal barrier for the $p$ wave gives rise to the formation of
the halo. Having a small $p$ wave component, the 6th orbital
($\Omega^\pi = 1/2^-$) contributes less to the halo though it is in
the continuum and the occupation probability is rather large. The
contribution of the 8th orbital ($\Omega^\pi = 1/2^+$) to the tail
of the density is even smaller because its main components are of
$d$ waves. The large centrifugal barrier of $f$ states hinders
strongly the spatial extension of the wave functions of the other
two continuum orbitals, i.e., the 5th ($\Omega^\pi = 7/2^-$) and the
7th ($\Omega^\pi = 3/2^-$).

\iffalse %

\begin{figure}[b]
\includegraphics[width=9cm]{fig4.eps}\hspace{2pc}%
\begin{minipage}[b]{14pc}
\caption{\label{fig4} (Color online) Decomposition of the neutron
density of $^{44}$Mg into spherical ($\lambda=0$), quadrupole
($\lambda=2$), and hexadecapole ($\lambda=4$) components for (a) the
core and (b) the halo. This figure is originally published in Ref.~\cite{Zhou2010}.}
\end{minipage}
\end{figure}

In Figure~\ref{fig4} the densities of the core and the halo are
decomposed into spherical, quadrupole, and hexadecapole components.
As is seen in Figure~\ref{fig4}a, the quadrupole component of the
core is positive, thus being consistent with the prolate shape of
$^{44}$Mg. However, for the halo, the quadrupole component has a
negative sign, which means that the halo has an oblate deformation.
The quadrupole moments of the neutron core and the halo are 160 and
$-$27 fm$^2$, respectively. This explains the decoupling between the
quadrupole deformations of the core and the halo as we have seen it
in Figs.~\ref{fig1}b and~\ref{fig1}c. There is also a noticeable
hexadecapole component in the density distribution of the halo.

\fi

The slightly oblate shape of the halo originates from the intrinsic
structure of the weakly bound and continuum orbitals. As is mentioned
above and shown in Figure~\ref{fig2}, the main Woods-Saxon components
of the two weakly bound orbitals, the 3rd ($\Omega^\pi = 1/2^-$) and
the 4th ($\Omega^\pi = 3/2^-$), are $p$ states. We know that the
angular distribution of $|Y_{10}(\theta,\phi)|^2 \propto
\cos^2\theta$ with a projection of the orbital angular momentum on
the symmetry axis $\Lambda=0$ is prolate and that of
$|Y_{1\pm1}(\theta,\phi)|^2 \propto \sin^2\theta$ with $\Lambda=1$ is
oblate. It turns out that in the 3rd ($\Omega^\pi = 1/2^-$) orbital,
both $\Lambda = 0$ and $\Lambda = 1$ components contribute and the
latter dominates. Therefore this orbital has a slightly oblate shape.
For the 4th ($\Omega^\pi = 3/2^-$) state, there is only the
$\Lambda=1$ component from the $p_{3/2}$ wave, an oblate shape is
also expected.

\section{Summary}

Neutron halo in deformed nuclei is investigated within a deformed
relativistic Hartree Bogoliubov model in a Woods-Saxon basis. In a
very neutron-rich deformed nucleus $^{44}$Mg a pronounced deformed
neutron halo is found. It is formed by several orbitals close to the
threshold. These orbitals have large components of low $\ell$-values
and feel therefore only a small centrifugal barrier. Although
$^{44}$Mg and its core are prolately deformed, the deformation of the
halo is slightly oblate. This implies a decoupling between the shapes
of the core and the halo. The mechanism is investigated by studying
the details of the neutron densities for core and halo, the single
particle levels in the canonical basis, and the decomposition of the
halo orbitals. We also studied the weakly-bound nuclei in Ne isotopes
and discussed the conditions for the occurence of a halo and the
shape decoupling~\cite{Zhou2010}. It is shown that the existence and
the deformation of a possible neutron halo depends essentially on the
quantum numbers of the main components of the single particle orbits
in the vicinity of the Fermi surface.

\ack This work has been supported in part by Natural Science
Foundation of China (10775004, 10705014, 10875157, and 10979066),
Major State Basic Research Development Program of China
(2007CB815000), Knowledge Innovation Project of Chinese Academy of
Sciences (KJCX3-SYW-N02 and KJCX2-YW-N32), by the Bundesministerium
f\"{u}r Bildung und Forschung (BMBF), Germany, under Project 06 MT
246, and by the DFG cluster of excellence \textquotedblleft Origin
and Structure of the Universe\textquotedblright\
(www.universe-cluster.de). The computation was supported by
Supercomputing Center, CNIC of CAS.

\section*{References}
%\begin{thebibliography}{9}
%\bibitem{iopartnum} IOP Publishing is to grateful Mark A Caprio, Center for Theoretical Physics, Yale University, for permission to include the {\tt iopart-num} \BibTeX package (version 2.0, December 21, 2006) with  this documentation. Updates and new releases of {\tt iopart-num} can be found on \verb"www.ctan.org" (CTAN).
%\end{thebibliography}

%\bibliographystyle{iopart-num-sgzhou}
%\bibliography{../../information/refs/JabRef/sgzhou}

%\end{thebibliography}

\providecommand{\newblock}{}

\end{document}